\documentclass[prb,twocolumn,preprintnumbers,amsmath,floatfix]{revtex4}
\usepackage{graphicx}

\begin{document}

\title{Microscopic model for transitions from Mott to spin-Peierls insulator in TiOCl}
\author{Yu-Zhong Zhang, Harald O. Jeschke, and Roser Valent\'\i}
\affiliation{Institut f\"ur Theoretische Physik,
Goethe-Universit\"at Frankfurt, Max-von-Laue-Stra{\ss}e 1, 60438
Frankfurt am Main, Germany}

\date{\today}

\begin{abstract}
On the basis of \textit{ab initio} density functional theory (DFT)
calculations, we derive the underlying microscopic model
Hamiltonian for TiOCl, a unique system that shows two consecutive
phase transitions from a Mott insulator to a spin-Peierls insulator
through a structurally incommensurate phase.
We show with our model that the presence of
magnetic frustration in TiOCl leads to a competition with the
spin-Peierls distortion, which results in the novel incommensurate
phase. In addition, our calculations indicate that the spin-Peierls
state is triggered by adiabatic phonons, which is essential for
understanding the nature of the phase transition.
\end{abstract}


\maketitle

\section{Introduction}
\label{sec:one}

Mott insulators are materials which owe their insulating properties
to strong electron correlations~\cite{Mott}. In practice, the Mott
state is unstable to residual interactions~\cite{Imada} and in one
dimension the dominant instability is the doubling of the unit cell
via the coupling of spins to phonons, so called the spin-Peierls
transition~\cite{Miller}.

With the discovery in the early nineties of the first inorganic
spin-Peierls material, CuGeO$_3$~\cite{Hase}, followed by
NaV$_2$O$_5$~\cite{NaV2O5} a large amount of work~\cite{Lemmens_03}
was devoted to the understanding of the complexity of this
instability when additional degrees of freedom like next nearest
neighbor interactions (CuGeO$_3$) or charge disproportionation
(NaV$_2$O$_5$) are involved. Moreover, commensurate to structurally
incommensurate transitions were observed in TTF-CuBDT~\cite{Keimer1}
and CuGeO$_3$~\cite{Keimer2,Lorenz} when a strong magnetic field is
applied.

Recently, the anomalous behavior of TiOCl~\cite{Seidel,Hoinkis,Lemmens_2} --
the third inorganic spin-Peierls compound with highest transition
temperature and largest lattice distortion yet -- has provoked a new
debate. Upon cooling, TiOCl undergoes two consecutive phase
transitions at $T_{c_2}=92$~K and $T_{c_1}=66$~K of second and first
order, respectively. These two temperatures separate a paramagnetic
Mott insulator at high temperatures ($T>T_{c_2}$) from a non-magnetic
dimerized spin-Peierls state at low temperatures ($T<T_{c_1}$). In the
intermediate region $T_{c_1}<T<T_{c_2}$, TiOCl shows structural
incommensurabilities ~\cite{Shaz,Krimmel,Rueckamp,Abel}. The
observation of such a phase without the need of applying a magnetic
field is unprecedented, making the study of TiOCl compelling since
this phase must be a result of the subtle competition between all
interactions present in the material.

\begin{figure}[h]
\includegraphics[width=0.43\textwidth]{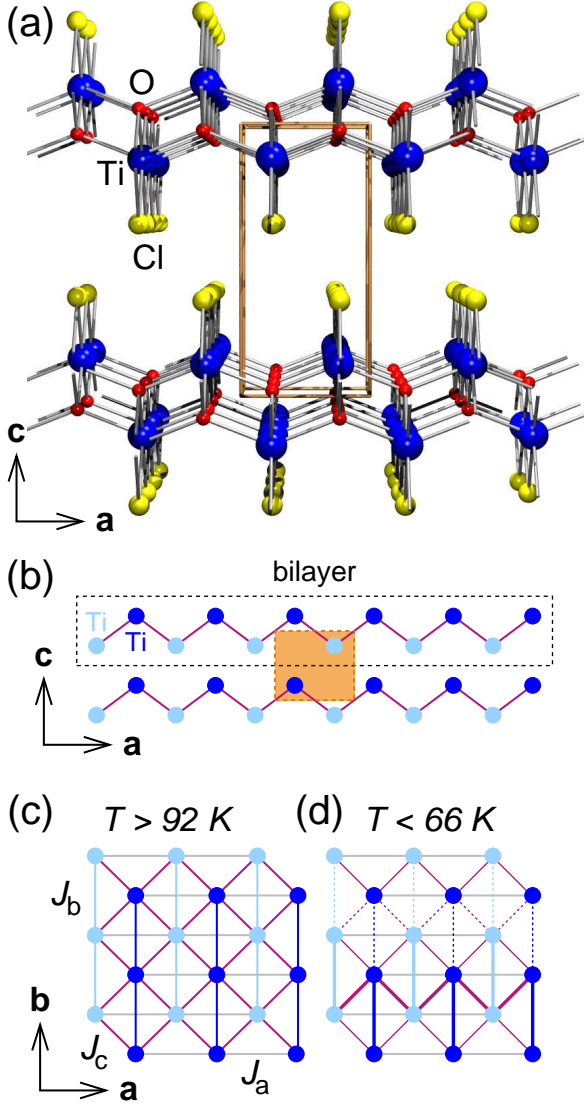}
\caption{(Color online) TiOCl structure and exchange interaction
diagrams. (a) View along $b$ direction of TiOCl. Here, Ti is shown
by blue ball (largest size), O red (smallest size) and Cl yellow
(intermediate size). (b)-(d) Schematic view of the structure,
showing only Ti ions. (b) Projection on the $ac$ plane. Blue
(Black)/light blue (light grey) circles denote Ti ions in
upper/lower layer. The unit cell is shown as a shaded region. (c),
(d) Projection of the bilayer onto the $ab$ plane.  Ti chains run
along $b$.  Note the triangular arrangement of Ti ions. Couplings
between Ti ions are shown as blue (vertical) ($J_b$), magenta
(diagonal) ($J_c$) and grey lines (horizontal) ($J_a$). The various
line thicknesses signal the strengths of Ti displacements for $T <
T_{c_1}$ (d) which are absent for $T>T_{c_2}$ (c).}
\label{fig:struct}
\end{figure}

TiOCl is also unconventional in the size of the distortion
$\eta=0.18$~{\AA} observed between neighboring Ti atoms along
$b$~\cite{Shaz,Krimmel} in the dimerized phase in contrast to
\textit{e.g.} polyacetylene where $\eta=0.08$~{\AA}~\cite {Yannoni}.
Moreover, the characteristic frequency of the spin-Peierls phonon
$\Omega_0\simeq27$~meV is comparable to the spin excitation gap
$\Delta \simeq 21$~meV associated with the static dimerization in
TiOCl for $T<T_{c_1}$~\cite{Abel}.  This indicates that the
anomalous spin-Peierls transition might be at the borderline between
adiabatic (the phonons behave classically) and non-adiabatic (the
quantum nature of the phonons is important for the phase transition)
behavior ~\cite{Bursill,Uhrig,Gros98}. Moreover, increasing interest
in applying pressure to TiOCl
~\cite{Kuntscher,Forthaus,Blanco-Canosa,Kuntscher1,Zhang} requires
decent understanding of the phase transitions at ambient pressure.

Looking at the structure of TiOCl, this system consists of double
layers of titanium (Ti) and oxygen (O) in the $ab$ plane separated
by chlorine (Cl) layers along the $c$ direction ~\cite
{Shaz,Krimmel,Seidel} (see~Fig.~\ref{fig:struct}~(a)). The Ti atoms
form chains along $b$ with intrachain coupling $J_b$ whose
projection on the $ab$ plane gives a triangular pattern
(Fig.~\ref{fig:struct}~(c) and (d)). It has been suggested in the
past~\cite{Shaz,Krimmel,Rueckamp} that possible frustrating
interchain interactions ($J_a$ within one layer and $J_c$ between
layers) could be responsible for the existence of the intermediate
structurally incommensurate phase with a temperature dependent
modulation wave vector
$\mathbf{q}=(q_1,0.5+q_2,0)$~\cite{Shaz,Krimmel,Abel}, with
$0<q_1,q_2\ll0.5$. But the mechanism driving the system to such a
phase has not been demonstrated.

In this paper, we will demonstrate that the behavior of TiOCl can be
described by a two-dimensional frustrated spin-Peierls model with
large magnetoelastic coupling. We show that magnetic
frustration~\cite{Ramirez} is responsible for the existence of a
structurally incommensurate phase at $ T_{c_1} < T < T_{c_2} $ . Our
results indicate that the spin-Peierls state is triggered by
adiabatic phonons. The model parameters in our proposed model are
calculated within \textit{ab initio} density functional theory (DFT)
by the Car-Parrinello projector augmented wave (CP-PAW)
method~\cite{Bloechl}.

The paper is organized as follows: Section \ref{sec:two} describes
the model we proposed and the details about how we determined the
model parameters by DFT methods. In Section \ref{sec:three} we
present our results of the model parameters and discuss the
consistencies between our results derived from the proposed model
and the experimental observations. The possible mechanism for the
structural incommensurability without external magnetic field is also
discussed and the nature of the phonons is  determined.
 Finally in Section \ref{sec:four} we present our conclusions.

\section{Model and method}
\label{sec:two}

The most general two-dimensional frustrated spin-Peierls model for
TiOCl can be written as follows:
\begin{align}
H&=H_a+H_b+H_c \label{eq:spin-Peierls} \\
H_a&=J_a\sum_{i,j}\mathbf{S}_{i,j}\cdot \mathbf{S}_{i+a,j}
\label{eq:spin-Peierlsa} \\
H_b&=J_b\sum_{i,j}\left( 1+\alpha _bd\xi _{ij,ij+b}\right)
\mathbf{S}%
_{i,j}\cdot \mathbf{S}_{i,j+b} \nonumber \\
&+\sum_{i,j}\frac{1}{2}K\left( d\xi _{ij,ij+b}\right)
^2+\sum_{i,j}\frac{P_{b,ij}^2}{2M_b} \label{eq:spin-Peierlsb} \\
H_c&=J_c\sum_{i,j}\left( 1+\alpha _cd\xi _{ij,i+\frac{a}{2}j\pm
\frac{b}{2}}\right) \mathbf{S}_{i,j}\cdot
\mathbf{S}_{i+\frac{a}{2},j\pm \frac{b}{2}}
\nonumber \\
&+\sum_{i,j}\frac{1}{2}K\left( d\xi _{ij,i+\frac{a}{2}j\pm
\frac{b}{2}}\right)^2+\sum_{i,j}\frac{P_{c,ij}^2}{2M_c},
\label{eq:spin-Peierlsc}
\end{align}
where $d\xi_{ij,kl}$ denote the displacements of the Ti ions in the
$ab$ plane $d\xi_{ij,kl}=\xi_{ij,kl}-\xi_{ij,kl}^0$. Here, $\xi
_{ij,kl}=\left| \mathbf{r}_{ij}-\mathbf{r}_{kl}\right| $ is the
distance between two Ti ions and
$\mathbf{r}_{ij}=\mathbf{r}_{ij}^0+\mathbf{u}_{ij}$ is the Ti
position for $T < T_{c_1}$ (Fig.~\ref {fig:struct}~(d)) obtained by
adding a displacement $\mathbf{u}_{ij}$ to the original Ti position
$\mathbf{r}_{ij}^0$ for $T > T_{c_2}$ (Fig.~\ref {fig:struct}~(c)).
$\alpha_b$ and $\alpha_c$ denote the spin-phonon couplings and $K$
is the elastic constant. $M_b$ and $M_c$ are the effective masses
obtained by weighted summation of the masses of Ti, O, and Cl atoms
and $P_{b,ij}$ and $P_{c,ij}$ denote the momentum at site $(i,j)$
along $b$ and $c$ respectively. Displacements along $a$ are not
included since they are absent in the low-symmetry
structure~\cite{Shaz,Krimmel}.

In order to obtain the model parameters from first principles
analysis, a fully relaxed $T>T_{c_2}$ crystal structure is needed.
Calculations with various functionals (generalized gradient
approximation (GGA) and GGA+U) with the CP-PAW method showed that
the lattice parameters as well as distances and angles from GGA+U
are more consistent with experimental results than those from GGA,
demonstrating the importance of including electron correlation
effects for the description of the lattice properties of
TiOCl~\cite{Pisani,Pisani1}, even if it is done at the simple level
of GGA+U. Moreover, our CP-PAW calculations are all carefully
checked to be fully converged with respect to the size of the basis
set.

\begin{figure}[h]
\includegraphics[width=0.47\textwidth]{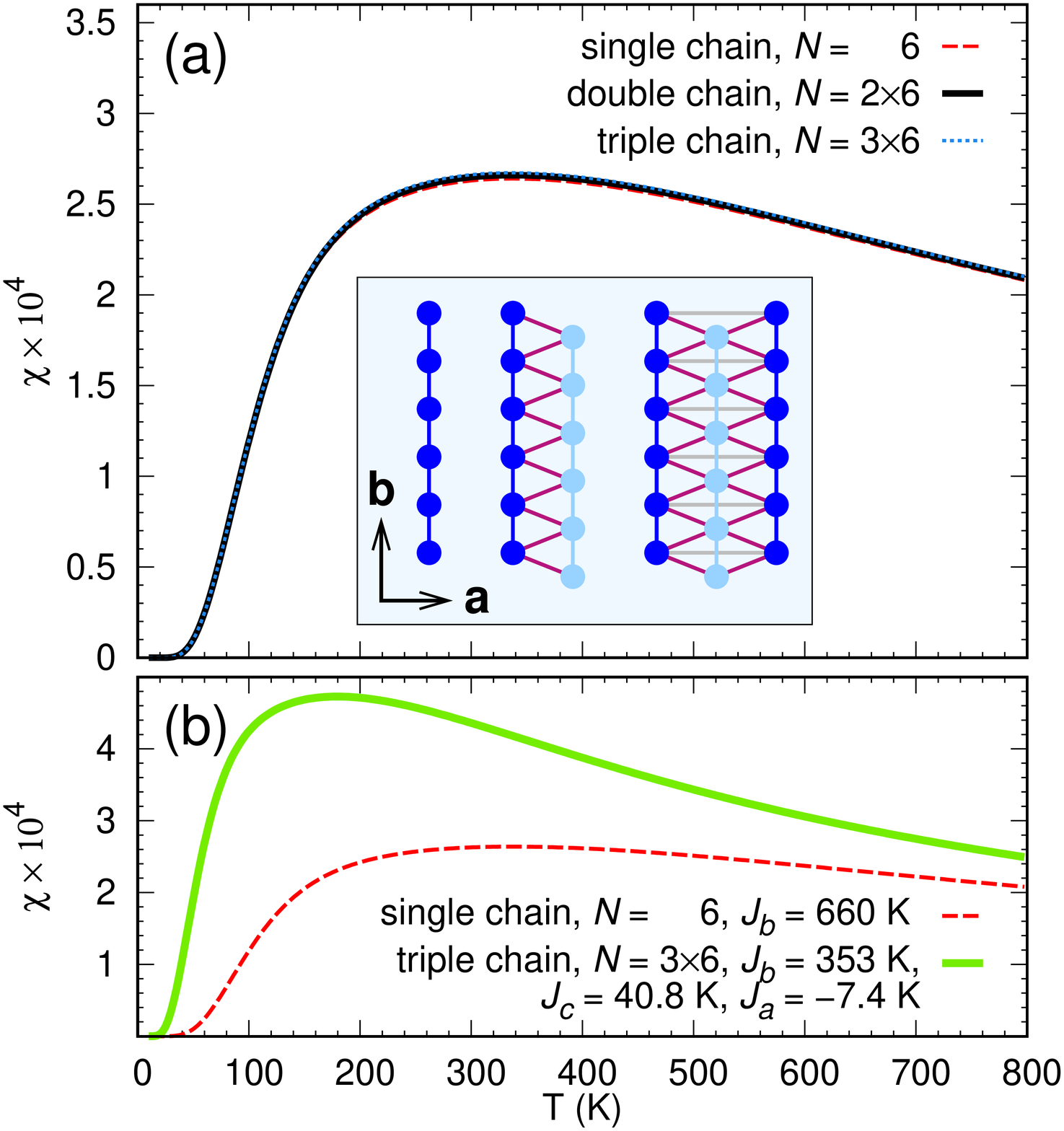}
\caption{(Color online) Exact diagonalization results for spin
susceptibility. (a) Susceptibility, calculated with the model
parameters determined at $U=1.65$~eV.  The 1-, 2-, and 3-chain
geometries are shown in the inset. Vertical/diagonal/horizontal
bonds denote the $J_{b}$/$J_{c}$/$J_{a}$ exchange coupling,
respectively. Blue (Black)/light blue (light grey) circles denote Ti
ions in upper/lower layer as Fig.~\ref{fig:struct}. (b)
Susceptibility calculated with the model parameters of
Ref.~\protect\onlinecite{Macovez} which clearly disagrees with the
susceptibility calculated for the experimentally determined coupling
$J_b=660$~K. } \label{fig:suscept}
\end{figure}

The magnetic exchange coupling constants $J_i = J_b, J_a, J_c$ were
obtained from total CP-PAW energy differences between the relevant
ferromagnetic (FM) and antiferromagnetic (AFM) Ti spin
configurations $E_\text{FM}-E_\text{AFM}$ by relating them to the
$J_i$ through mapping to the classical Heisenberg model. The
determination of the spin-phonon couplings $\alpha_b$ and $\alpha_c$
was more elaborate. Out of the phonon spectrum of TiOCl for the
high-symmetry ($T>T_{c_2}$) and the low-symmetry (dimerized)
($T<T_{c_1}$) phase~\cite{Pisani}, the spin-Peierls phonon was
identified and found to be degenerate in energy (A$_g$ and B$_u$).
The displacement vector for the Ti atoms
$\mathbf{u}_{i{\genfrac{}{}{0pt}{}{j}{j+1}}}=(0.0,\mp 0.0223,\pm
0.0403)\delta$, corresponding to the spin-Peierls phonon, was then
calculated by diagonalizing the dynamical matrix, where $\delta$ is
the order parameter, with $\delta =2$~corresponding to the
displacements of the Ti atoms in the dimerized phase. We obtained
$\alpha_c$ by calculating the difference in energy between the FM
and a spin $S=2$ AFM spin arrangement of the unit cell doubled in
both $b$ and $c$ directions and distorted according to the
displacement vectors $\mathbf{u}_{ij}$ for $\delta = 2$. $\alpha_b$
was calculated analogously. We performed the calculations for
various $U$ values and found that $J_{c}$ and $J_{a}$ are always FM
and relatively small compared to the exchange coupling $J_{b}$; the
latter is always AFM and decreases monotonously as $U$ increases,
roughly fulfilling the relation $J\sim 1/U$. We have also performed
LAPW calculations with the WIEN2k~\cite{Blaha} code to check these
results and we obtain good agreement between both
methods.

The elastic constant $K$ was calculated by the expression $E\left(
\delta \right) =E\left( 0\right) +0.0072NK\delta^2/2$ where $N$ is
the number of Ti ions in the unit cell and the numerical factor in
the $\delta^2$ term arises from the relationship between $\delta$
and $d\xi$ (see equation~\ref{eq:spin-Peierlsc}). $E\left( \delta
\right) $ is the CP-PAW ground state energy obtained for distorted
lattices according to the spin-Peierls $\mathbf{u}_{ij}$ vectors
given above. $K$ was found to be almost independent of the choice of
$U$.

\section{Results and discussions}
\label{sec:three}

For the moderate value $U=1.65$~eV~\cite{comment_J}, we obtain the
model parameters $J_b=660.1$~K, $J_c=-16.7$~K, $J_a=-10.5$~K,
$\alpha_b=1.73$~{\AA}$^{-1}$, $\alpha_c=8.86$~{\AA}$^{-1}$ and
$K=0.292$~eV/{\AA}$^2$. The reliability of these parameters is
demonstrated by comparing the spin susceptibilities among single-,
double- and triple-chain Heisenberg models calculated by exact
diagonalization (see Fig.~\ref{fig:suscept}~(a)). The behavior of
the spin susceptibility at moderate to high temperatures remains
almost unchanged when interchain couplings $J_c$ and $J_a$ are
gradually introduced from single- to triple-chain models. This
insensitivity of the susceptibility to the interchain couplings
explains why initially TiOCl was described as a good realization of
a one-dimensional spin 1/2 system~\cite{Seidel} since the
experimental data could be perfectly fitted to a one-dimensional
spin 1/2 Heisenberg chain. Our results show that the experimental
susceptibility is well described by a frustrated model. In contrast,
a recent estimate of the magnetic interactions given by Macovez
\textit{et al.}~\cite{Macovez} for TiOCl with $J_b=353$~K,
$J_a=-7.4$~K and $J_c=40.8$~K shows in our ED calculations
considerable deviations from the experimental susceptibility as
shown in Fig.~\ref{fig:suscept}~(b).

\begin{figure}[h]
\includegraphics[width=0.46\textwidth]{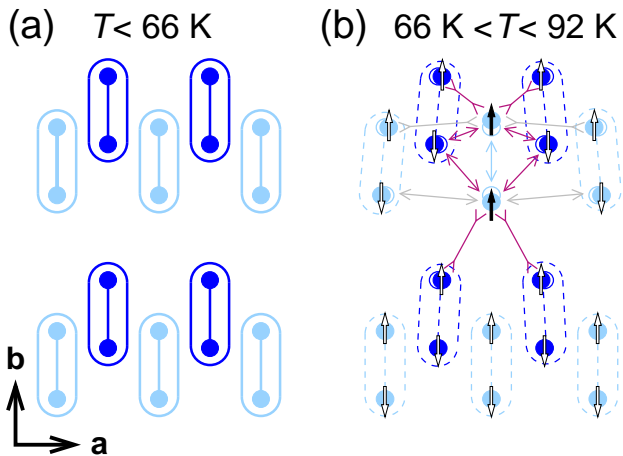}
\caption{(Color online) Explanation of incommensurate structure.
Here, blue (black)/light blue (light grey) circles denote Ti ions in
upper/lower layer as Fig.~\ref{fig:struct}. (a) Dimerized structure
at temperatures below $T_{c_1}=66$~K. The antiferromagnetic
interactions and spin-phonon coupling along $b$ lead to the
formation of stable singlets. (b) Above $T_{c_1}=66$~K, thermal
fluctuations lead to a competition between antiferromagnetic
interactions along $b$ and frustrating ferromagnetic interactions
along $a$ and $c$. We illustrate the effect on the atomic positions
by showing the result of a spin flip in the central Ti dimer (black
solid spins). The resulting attractive (inward arrows) or repulsive
(outward arrows) interactions (gray (light grey) line $J_a$ and
magenta (dark grey) $J_c$) lead to modulations of the Ti atom
positions along $a$ and $b$ directions, explaining the formation of
an incommensurate structure. } \label{fig:incomm}
\end{figure}

The obtained values of $J_a$ and $J_c$ are the key for understanding
the two consecutive phase transitions in TiOCl. The cooperation of
$J_a$ and $J_c$ ($2J_c+J_a=-43.9$K) in competition with $J_b$ on a
Ti site is in the range of energies where the incommensurability
happens in the system (between $T_{c_1}=66~$K and $T_{c_1}=92$~K).
Starting from the non-magnetic dimerized phase, this ferromagnetic
interaction together with thermal fluctuations suppresses the
formation of dimer states in favor of formation of spin 1/2
solitons leading to a structurally incommensurate phase. In
Fig.~\ref{fig:incomm} we present an illustration of the process. As
temperature increases, the ferromagnetic coupling breaks the central
spin zero dimer into a triplet state, and the neighboring spins get
partially polarized. These spins feel the attraction (repulsion) of
the central Ti $S=1/2$ spins depending on whether their interaction
is favorable (unfavorable) according to $J_a$, $J_c$ and $J_b$.
Consequently, helped by the thermal fluctuations, the neighboring
Ti atoms will displace towards (away from) the central spins (see
Fig.~\ref{fig:incomm}~(b)). This displacement will propagate to the
next nearest neighbor atoms so that a sinusoidal atomic
displacement (soliton) over many unit cells along the $a$ and the
$b$ direction will appear, which defines the incommensurability
along $a$ and $b$ observed experimentally.

The couplings in TiOCl reveal still further important features of
the spin-Peierls phase transition. Since $J_b\gg J_a,J_c$, we can
define a dimensionless magnetoelastic coupling constant, which
uniquely determines the properties of the system, as
$\lambda=J_b\alpha_b^2/K$ with $\lambda_\text{DFT}=0.58$ as
calculated from our model parameters for TiOCl. This value is
significantly larger than $\lambda=0.32$ for
polyacetylene~\cite{Yannoni}. With $\lambda$ we can now tackle the
question whether the spin-Peierls transition is adiabatic or
non-adiabatic.

\begin{figure}[h]
\includegraphics[angle=-90,width=0.47\textwidth]{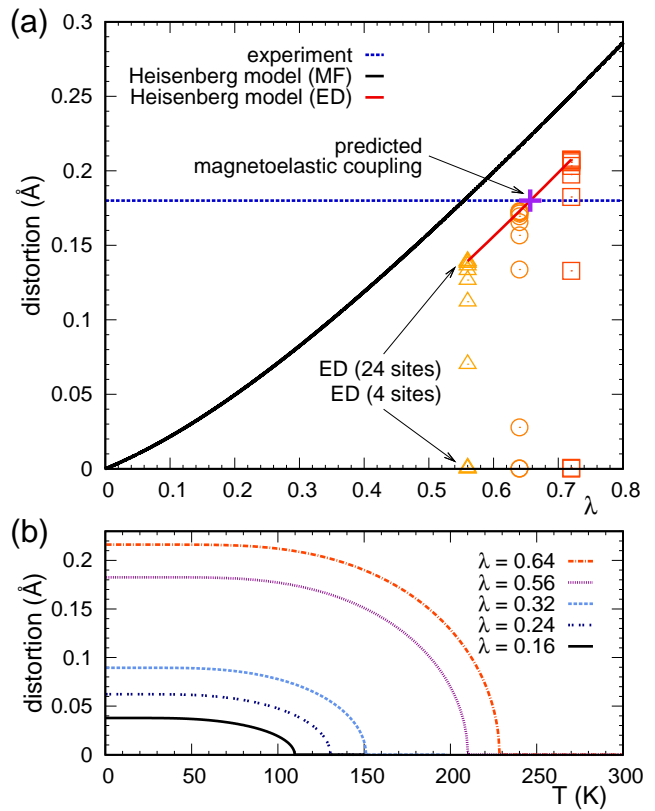}
\caption{(Color online) Predictive power of calculated model
parameters. (a) Lattice distortion $\eta$ as a function of the
magnetoelastic coupling $\lambda$ obtained from various approaches:
Heisenberg model in mean field approximation (black curve) and exact
diagonalization calculations for systems up to 24 sites (symbols)
for $\lambda=0.56$ (triangles), $0.64$ (circles), and $0.72$
(squares).  Note that sizes of $N=24$ are large enough to reach
converged results. The horizontal dashed line denotes the
experimental $\eta$.  (b) Lattice distortion $\eta$ as a function of
temperature for various $\lambda$ obtained for $H_b$ in the mean
field approximation. The second order nature of the $T_{c_2}$ phase
transition is evident from the curves. } \label{fig:distort}
\end{figure}

We analyze $H_b$ (equation~\eqref{eq:spin-Peierlsb}) in the
adiabatic limit by (i) using the mean field (MF) approximation after
applying a Jordan-Wigner transformation and by (ii) performing exact
diagonalization (ED) calculations for chains with up to $24$ spins
at $T=0$. In Fig.~\ref{fig:distort}~(a), we show the distortion of
the lattice $\eta$ as a function of the coupling constant $\lambda $
for the two calculations. The MF approximation predicts a
$\lambda_\text{MF}\approx0.56$ to induce the experimental
distortion. Note that this $\lambda_\text{MF}$ is very close to our
\textit{ab
  initio} determined value $\lambda_\text{DFT}$.  ED results
qualitatively agree with the MF results and predict a
$\lambda_\text{ED}\approx0.64$ to describe the experimental
distortion. This agreement between $\lambda_\text{DFT}$ and
$\lambda_\text{MF}$ or $\lambda_\text{ED}$ demonstrates the high
degree of consistency in our calculations. In
Fig.~\ref{fig:distort}~(b), we show the temperature dependence of
the phase transition within the MF approach for different $\lambda$
values. We observe that the spin-Peierls phase transition is of
second order and the critical temperature $T_\text{SP}$ lies between
$100$ and $230$~K depending on the choice of $\lambda$, which is not
far away from the experimental $T_{c_2}$.  As is well known,
inclusion of interchain couplings and strong correlation in the
calculation further reduces $T_\text{SP}$ to the experimental value
$T_{c_2}$.

We now consider $H_b$ (equation~\eqref{eq:spin-Peierlsb}) in the
non-adiabatic limit. After quantizing the lattice degrees of freedom
by introducing the phonon creation and annihilation operators and
integrating out the phonon degrees of freedom, an effective
$J_1$-$J_2$ model is obtained, where
$J_2=\left(J_b\alpha_b\right)^2/4M_b\Omega_0^2\approx0.036J_b$ and
$J_1=J_b+2J_2\approx 1.07J_b$ ($\Omega_0=2\sqrt{K/M_b}$ is the bare
phonon frequency). Since the ratio $J_2/J_1<0.24$~\cite{Eggert},
non-adiabaticity will not affect the dimerization in TiOCl (in
contrast to the situation in CuGeO$_3$~\cite{Uhrig,Gros98}). In fact,
with our calculated model parameters, we find that
$\Omega_0/T_\text{SP}\approx2$ fulfills the condition
$\Omega_0/T_\text{SP}<2.2$~\cite{Gros98} for the occurrence of a soft
phonon. Out of these two limiting calculations, we conclude that the
main features of TiOCl can be well described in the adiabatic limit,
in agreement with recent experiments~\cite{Abel}.

\section{Conclusions}
\label{sec:four}

The presented work shows that TiOCl is described by a frustrated
spin-Peierls model on an anisotropic triangular lattice with
intrachain AFM and interchain FM interactions and large
magnetoelastic coupling. This is in contrast with previous
suggestions on the magnetic interactions.  With this model we can
account for the Mott to spin-Peierls insulator phase transition in
TiOCl and for the existence of intermediate zero-field structural
incommensurabilities.  This work also shows that the combination of
{\it ab initio} DFT calculations with effective models is a powerful
tool to describe the microscopics of this Mott insulator.

\section{Acknowledgments}

We acknowledge useful discussions with L. Pisani, C. Gros, F.
Essler, M. Mostovoy, D. Khomskii, C. Walther, P. Bl\"ochl, M. Sing
and R. Claessen. We thank the Deutsche Forschungsgemeinschaft for
financial support through the TRR/SFB~49 and Emmy Noether programs
and we gratefully acknowledge support by the Frankfurt Center for
Scientific Computing.

\appendix
\section{Mean-field model calculations}

By means of Jordan-Wigner transformation to $H_b$
(equation~\eqref{eq:spin-Peierlsb}), a spin-less fermion model is
obtained,
\begin{align}
\frac{H_b^{\prime }}{J_b}&=\sum_i\left( 1+\left( -1\right) ^i\Delta
\right)
\left[\frac{1}{2}\left( c_i^{\dagger }c_{i+1}+h.c.\right)\right.  \nonumber \\
&\left.+\left( n_i-\frac{1}{2}\right) \left(
n_{i+1}-\frac{1}{2}\right)\right] +\sum_i\frac{1}{2}\frac{\Delta
^2}{\lambda}\,,
\end{align}
where $\Delta =\alpha_b\xi_{ij,ij+b} $, $c_i^{\dagger }\left(
  c_i\right) $ is the creation (annihilation) operator of an electron
and $n_i=c_i^{\dagger }c_i$. Applying the Hartree-Fock approximation
to the interaction part and with the help of Fourier transformation
and unitary transformation, three coupled self-consistent equations
can be obtained from $\frac{\partial F}{\partial \Delta }\bigl|_\mu
=0$, $\frac{\partial F}{\partial b_0}\bigl|_\mu =0$ and
$\frac{\partial F}{\partial \delta b}\bigl|_\mu =0$ where $F$ is the
free energy, and the chemical potential $\mu $ is adjusted to yield
the correct filling, {\it i.e.}, $N_e=\frac{1}{\beta}
\frac{\partial}{\partial \mu }\ln \Xi $. Here, $\beta =1/k_BT$ and
$\Xi $ is the grand canonical partition function
\begin{equation}
\ln \Xi =\sum_{k,\pm }\ln \left[ 1+e^{-\beta \left( \epsilon
_\text{HF}^{\pm }\left( k\right) -\mu \right) }\right] ,
\end{equation}
where the band structure $\epsilon_\text{HF}^{\pm }\left( k\right)
=\pm \sqrt{E\left( k\right) }$ and
\begin{align}
E\left( k\right) &=\left( 1-\Delta \right) ^2\left(
\frac{1}{2}-b_0+\delta
b\right) ^2 \nonumber \\
&+\left( 1+\Delta \right) ^2\left( \frac{1}{2}-b_0-\delta b\right) ^2 \\
&+\left( 1-\Delta ^2\right) \left( \left( \frac{1}{2}-b_0\right)
^2-\left( \delta b\right) ^2\right) 2\cos k. \nonumber
\end{align}
$b_0$ and $\delta b$ is defined as $\left( \left\langle c_i^{\dagger
}c_{i+1}\right\rangle +\left\langle c_{i+1}^{\dagger
}c_{i+2}\right\rangle \right) /2$ and $\left( \left\langle
c_i^{\dagger }c_{i+1}\right\rangle -\left\langle c_{i+1}^{\dagger
}c_{i+2}\right\rangle \right) /2$, respectively. Finally, the
self-consistent equations read
\begin{align}
\Delta &=\frac{\lambda}{2}\left[\frac{1}{N_e}\sum_k\left\{ f\left(
\epsilon_\text{HF}^{-}\left( k\right) \right) -f\left( \epsilon
_\text{HF}^{+}\left( k\right)
\right) \right\}\times \right.\nonumber \\
&\left.\mspace{110mu}\times \frac{1}{2\sqrt{E\left( k\right) }}\frac{\partial E\left( k\right) }{%
\partial \Delta }-4b_0\delta b\right] \\
b_0&=\frac{1}{4}\left[\frac{1}{N_e}\sum_k\left\{ f\left( \epsilon
_\text{HF}^{-}\left( k\right) \right) -f\left(
\epsilon_\text{HF}^{+}\left( k\right)
\right) \right\}\times  \right.\nonumber \\
&\left.\mspace{110mu}\times \frac{1}{2\sqrt{E\left( k\right) }}\frac{\partial E\left( k\right) }{%
\partial b_0}-4\Delta \delta b\right] \\
\delta b&=\frac{1}{4}\left[\frac{1}{N_e}\sum_k\left\{ f\left(
\epsilon_\text{HF}^{-}\left( k\right) \right) -f\left( \epsilon
_\text{HF}^{+}\left( k\right)
\right) \right\}\times  \right.\nonumber \\
&\left.\mspace{110mu}\times \frac{1}{2\sqrt{E\left( k\right) }}\frac{\partial E\left( k\right) }{%
\partial \delta b}-4\Delta b_0\right]
\end{align}
where $f\left( \epsilon \right) $ is the Fermi distribution
function.


\begin{thebibliography}{99}
\bibitem{comment_J}  By comparing the DOS and the gap, we note that the
$U$ values used in the CP-PAW  ($U = 1.65$~eV in our case)
correspond  to larger $U$ values in Wien2k ($U=3.3$~eV) due to the
different implementations of GGA+U in these two approaches. For
instance, we obtained within LAPW for $U=3.3$~eV: $J_b=613.4$~K,
$J_c=-16.1$~K, $J_a=-7.3$~K which compare very well with the CP-PAW
results for $U=1.65$~eV.
\end{thebibliography}
\end{document}